\documentclass[conference]{IEEEtran}
\usepackage{amsmath,amssymb,amsfonts}
\usepackage{graphicx}
\usepackage[dvipsnames]{xcolor}
\usepackage[doi=false,isbn=false,url=false,eprint=false,style = ieee,minbibnames=6,maxbibnames=6,citestyle=numeric-comp]{biblatex}
\usepackage{bm}
\ifCLASSOPTIONcompsoc
\usepackage[caption=false,font=normalsize,labelfon
t=sf,textfont=sf]{subfig}
\else
\usepackage[caption=false,font=footnotesize,listofformat=subsimple,labelformat=simple]{subfig}
\fi

\usepackage{stfloats}

\usepackage{enumitem}
\usepackage{tabularx}
\usepackage[normalem]{ulem}

\usepackage[scr=boondox]{mathalfa}

\AtBeginBibliography{\footnotesize}
\AtEveryBibitem{\clearfield{day}}

\newcommand{\matr}[1]{\mathbf{#1}}

\newcommand{\X}{\mathcal{X}}
\newcommand{\E}{\mathcal{E}(\X)}
\newcommand{\y}{\mathscr{y}} % set of measurements
\newcommand{\Y}{\mathscr{Y}} % set of random variables
\newcommand{\fnij}[1]{#1_{ij}}

\newcommand{\dij}{\fnij{d}^{\X}}

\newcommand{\measijm}{\fnij{y}^{[m]}}

\usepackage[
  colorlinks=true,
  linkcolor=Black,
  citecolor=Black,
  filecolor=black,
  urlcolor=blue,
]{hyperref}
\usepackage{booktabs}

\title{Mismatched Estimation\\ in the Distance Geometry Problem}
\author{\IEEEauthorblockN{Mahmoud Abdelkhalek, Dror Baron, and Chau-Wai Wong}
\IEEEauthorblockA{Electrical and Computer Engineering, North Carolina State University\\Email: \{maabdelk, barondror, chauwai.wong\}@ncsu.edu}}

\begin{document}
\maketitle
\begin{abstract}
We investigate mismatched estimation in the context of the distance geometry problem (DGP). In the DGP, for a set of points, we are given noisy measurements of pairwise distances between the points, and our objective is to determine the geometric locations of the points. A common approach to deal with noisy measurements of pairwise distances is to compute least-squares estimates of the locations of the points. However, these least-squares estimates are likely to be suboptimal, because they do not necessarily maximize the correct likelihood function. In this paper, we argue that more accurate estimates can be obtained when an estimation procedure using the correct likelihood function of noisy measurements is performed. Our numerical results demonstrate that least-squares estimates can be suboptimal by several dB.
\end{abstract}
\section{Introduction} \label{sec:intro}

The distance geometry problem (DGP) \cite{liberti_distance_2017-1,voller_distance_2013,crippen_distance_1988} aims to determine the locations of points in Euclidean space given measurements of the distances between these points. The DGP arises in many important applications. In wireless networks with moving devices in two-dimensional (2D) planes, the positions of these devices can be estimated by solving the DGP~\cite{eren_rigidity_2004}. In computational biology, solving the DGP can allow us to determine the locations of atoms in molecules \cite{havel_evaluation_1991}. In robotics, a simple robotic movement can be modeled as a pair of rigid bars rotating along a joint. Whether these bars can flex along a joint to reach a specific point in 3D space can be determined by solving the DGP \cite{liberti_open_2018}.

The main challenge for the DGP is that pairwise distance measurements are noisy \cite{liberti_distance_2017-1}. Because of the noise, it is common to find a set of locations of points that minimizes the sum of squared errors (SSE) between the pairwise distances of a target structure and noisy measurements of these pairwise distances \cite{lavor_overview_2009}. When the noisy measurements of pairwise distances are Gaussian distributed, minimizing the SSE function yields a maximum likelihood (ML) estimate of a structure of points. However, in practice, it is unlikely that the measurements are exactly Gaussian distributed, and so the SSE function is unlikely to yield an ML estimate. In this work, we will provide better estimation quality by performing ML estimation using the correct noise distribution. We call our approach {\em matched DGP estimation}. In contrast, much of the prior art performed {\em mismatched} DGP estimation (details below).

\begin{figure}[t]
\centering
\includegraphics[width=0.8\linewidth]{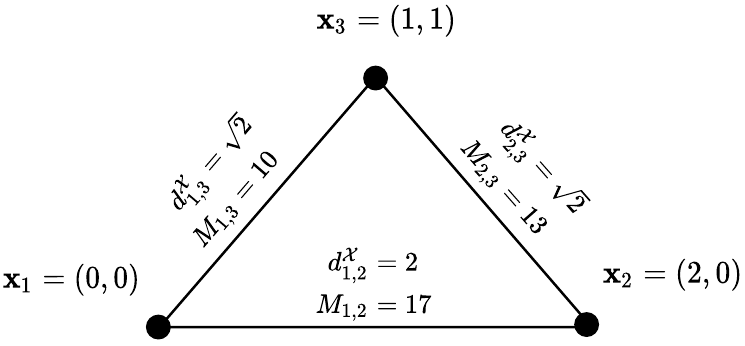}
\caption{Example structure $\X$ comprised of 3 points, $\mathbf{x}_1,\mathbf{x}_2,\mathbf{x}_3 \in \mathbb R^2$.}
\label{fig:tri_example}
\end{figure}

\textbf{Problem formulation.} Let $\X = \{\matr{x}_1,\matr{x}_2,\dots,\matr{x}_N\}$ be an unobserved set of locations of $N$ points in $K$-dimensional Euclidean space $\mathbb{R}^K$. Additionally, let $\E = \{(i,j) \mid (i,j) \in \{1,\dots,N\}^2, i < j\}$ be the set of edges connecting every pair of points in $\X$. Note that $|\E| = \frac{1}{2}N(N-1)$, where $|\cdot|$ denotes the cardinality of a set. We define the length $\dij$ of an edge $(i,j) \in \E$ as the $\ell_2$ norm of $\mathbf{x}_i - \mathbf{x}_j$, $\dij = \left\| \mathbf{x}_i - \mathbf{x}_j \right\|_2$. We define $\fnij{M} \in \mathbb{N}$, where $\mathbb N$ is the set of nonnegative integers, as the number of noisy measurements of the length of edge $(i,j) \in \E$. An example structure $\X$ is shown in \autoref{fig:tri_example}. In the DGP, our objective is to determine $\X$ given noisy measurements of lengths of edges in $\E$.

\textbf{Estimating the locations of the points.} We estimate $\X$ using the method of ML. Let the random variable $\fnij{Y}^{[m]} \in \mathbb R^+$ represent the $m$th noisy measurement of the length of edge $(i,j)$, where a realization of $\fnij{Y}^{[m]}$ is denoted by $\measijm$. For example, $y_{1,2}^{[8]}$ is the $8$th measurement of the length of edge $(1,2)$. We assume that $\fnij{M} = M$ for every $(i,j) \in \E$. Let the set of all $\fnij{Y}^{[m]}$ be, $\Y = \bigcup_{(i,j) \in \E} \fnij{\Y}$, where $\fnij{\Y} = \left\{\fnij{Y}^{[m]}\right\}_{m=1}^M$. The sets $\y$ and $\fnij{\y}$ are defined similarly for $\measijm$. Our objective is to derive an expression for the likelihood function $f_{\Y} (\y \mid d_{\E})$, where $d_{\E} = \left\{\dij\right\}_{(i,j) \in \E}$.

Suppose that the measurements for edge $(i_1,j_1)$ are independent of the measurements for edge $(i_2,j_2)$ for every $(i_1,j_1),(i_2,j_2) \in \E$, and that $\mathbb{E}\left[\fnij{Y}^{[m]}\right] = \dij$ for every $m = 1,\dots,M$. Under these assumptions, the likelihood can be expressed as, $f_{\Y} (\y \mid d_{\E}) = \prod_{(i,j) \in \E} f_{\fnij{\Y}}\left(\fnij{\y} \bigm| \dij\right)$. Additionally, suppose that all measurements associated with edge $(i,j)$ are independent and identically distributed (i.i.d.). The likelihood can be further simplified,
\begin{equation} \label{eq:ll}
f_{\Y} \left(\y \mid d_{\E}\right) = \prod_{(i,j) \in \E} \prod_{m = 1}^{M} f_{\fnij{Y}^{[m]}}\left(\measijm \Bigm| \dij\right),
\end{equation}
where $f_{\fnij{Y}^{[m]}}\left(\measijm \Bigm| \dij\right)$ is the \textit{true} marginal pdf of $\fnij{Y}^{[m]}$. In the rest of the paper, we will use $f_{\fnij{Y}^{[m]}}\left(\measijm \Bigm| \dij,\fnij{\bm{\theta}}\right)$, where $\fnij{\bm{\theta}}$ is a vector of shape parameters, to denote the marginal pdf for $\fnij{Y}^{[m]}$. When a single shape parameter describes this marginal pdf, the scalar version $\fnij{\theta}$ will be used instead.

Given a set of noisy edge length measurements $\y$, an ML estimate of $\X$ can be obtained by maximizing \eqref{eq:ll}. This case will be referred to as matched DGP estimation. In practice, we often do not know what parametric form the marginal pdf of $\fnij{Y}^{[m]}$ takes. In this case, we would modify~ \!\eqref{eq:ll} by replacing the true marginal pdf of $\fnij{Y}^{[m]}$ with an \textit{assumed} version, $f_{\fnij{Y}^{[m]}}^*\left(\measijm \Bigm| \dij\right)$. If $f_{\fnij{Y}^{[m]}}^*\left(\measijm \Bigm| \dij\right) \neq f_{\fnij{Y}^{[m]}}\left(\measijm \Bigm| \dij\right)$, then we refer to the modified version of the likelihood as the \textit{mismatched likelihood}. Moreover, an estimate of $\X$ that is obtained by maximizing the mismatched likelihood will not necessarily be an ML estimate. We refer to this case as mismatched DGP estimation. We are interested in the difference in quality of estimates of $\X$ in the matched and mismatched cases.

\textbf{Limitations of least-squares.} Suppose that $\fnij{Y}^{[m]}$ follows a Gaussian distribution with mean $\dij$ and variance $\fnij{\theta}$, namely, \[f_{\fnij{Y}^{[m]}}\left(\measijm \Bigm| \dij, \fnij{\theta}\right) = \frac{1}{\sqrt{2\pi\fnij{\theta}}} \exp\left(-\frac{(\measijm-\dij)^2}{2\fnij{\theta}}\right).\] Then, maximizing the likelihood $f_{\Y} \left(\y \mid d_{\E}\right)$ with respect to (w.r.t.) $\X$ is equivalent to minimizing the SSE w.r.t. $\X$. Several approaches to solve the DGP in the case of noisy measurements use variants of the SSE function \cite{crippen_distance_1988,hendrickson_molecule_1995,more_global_1997-2}. A survey of prior art appears in \autoref{sec:related_work}.

The aforementioned equivalence between minimizing the SSE function and maximizing the likelihood derived in \eqref{eq:ll} is only true when $\fnij{Y}^{[m]}$ is Gaussian. In practice, depending on the measurement technique, it is unlikely that the noisy measurement $\fnij{Y}^{[m]}$ is Gaussian distributed. Therefore, it is unlikely that minimizing the SSE function will yield an ML estimate of a structure.

\textbf{Contributions.} In this paper, we empirically show that, compared to mismatched DGP estimation, matched DGP estimation can offer an improvement of several dB. Moreover, our contribution is that, to the best of our knowledge, and in the context of the DGP, we are the first to focus on ML estimation beyond Gaussian-distributed measurements. The rest of the paper is structured as follows: we review the prior art in \autoref{sec:related_work}. The different marginal pdfs of $\fnij{Y}^{[m]}$ that will be considered, and how estimates of $\X$ will be evaluated, are discussed in \autoref{sec:proposed_method}. Experimental results are presented in \autoref{sec:results}. Finally, we summarize our results and describe future work in \autoref{sec:conclusion}\!.
\section{Related Work} \label{sec:related_work}

One of the earliest discussions of cost functions related to the solution of the DGP was due to Crippen and Havel \cite{crippen_distance_1988}. In their work, they used a variant of the SSE function, which we refer to as SSE-CH, to estimate molecular structures with physical constraints. Although others have empirically shown that SSE-CH yields acceptable results \cite{crippen_distance_1988}, the resulting estimate of a structure need not be an ML estimate.

Subsequent papers implement modifications to the SSE-CH cost function. For example, Mor\'e and Wu \cite{more_global_1997-2} adopted a continuation approach: the SSE-CH cost function is repeatedly smoothed via a convolution with a Gaussian kernel. After each successive smoothing stage, a minimizer of the smoothed SSE-CH is obtained. Eventually, the minimizers are traced back to the original function \cite[p. 827]{more_global_1997-2}. Although this approach yielded promising results, the underlying cost function that is smoothed is still the SSE function. Moreover, there is no guarantee that the resulting estimate of $\X$ is an ML estimate.

Souza et al. \cite{souza_hyperbolic_2011} attempt to make SSE-CH differentiable by approximating certain parts using hyperbolic functions. Souza et al. \cite{souza_solving_2013} made further improvements by applying these approximations to the \textit{geometric buildup} algorithm proposed by Luo and Wu \cite{luo_least-squares_2011}. However, the underlying cost function is still a variant of the SSE function, and so there is no guarantee of obtaining ML estimates. For a detailed review of approaches to solve the DGP, see \cite{lavor_overview_2009}. More recently, Gopalkrishnan et al. \cite{gopalkrishnan_dna_2020-1} optimize a variation of SSE-CH to reconstruct DNA molecules from pairwise distance measurements. Giamou et al. \cite{giamou_convex_2022} also optimize a variation of SSE-CH to determine the joint angles of a robot given the robot's forward kinematics and desired poses. However, because variations of SSE-CH are used, it is unlikely that estimates obtained in these applications are ML estimates.

\section{Noise Distributions and Evaluation} \label{sec:proposed_method}

We will consider different forms of the marginal pdf of $\fnij{Y}^{[m]}$, as included in \eqref{eq:ll}, that result in a likelihood $f_{\Y} \left(\y \mid d_{\E}\right)$ that, when maximized w.r.t. $\X$, yields ML estimates. It was previously shown that if $\fnij{Y}^{[m]}$ follows a Gaussian distribution, then minimizing the SSE function is equivalent to maximizing the likelihood in \eqref{eq:ll}. We consider the cases when $\fnij{Y}^{[m]}$ follows either a Laplace or non-standardized Student's $t$ (NSST) distribution.

If $\fnij{Y}^{[m]}$ follows a Laplace distribution with mean $\dij$ and variance $2\fnij{\theta}^2$, then
\begin{equation} \label{eq:laplace_pdf}
f_{\fnij{Y}^{[m]}}\left(\measijm \Bigm| \dij, \fnij{\theta}\right) = \frac{1}{2\fnij{\theta}} e^{-\left\lvert \measijm-\dij\right\rvert \big/ \fnij{\theta}}.
\end{equation}
If $\fnij{Y}^{[m]}$ follows an NSST distribution with mean $\dij$ and variance $\fnij{b}^2\fnij{\nu}/(\fnij{\nu}-2)$, assuming $\fnij{\nu} > 2$, then an expression for the marginal pdf of $\fnij{Y}^{[m]}$ can be derived as follows. Let $W$ follow a Student's $t$ distribution with parameter $\fnij{\nu}$, which is a positive integer. The pdf of $W$ is
\begin{equation*}
f_W(w \mid \fnij{\nu}) = \frac{\Gamma\left((\fnij{\nu} + 1)/2\right)}{\Gamma\left(\fnij{\nu}/2\right)\sqrt{\fnij{\nu} \pi}} \left(1 + (w^2 / \fnij{\nu})\right)^{-(\fnij{\nu} + 1)/2},
\end{equation*}
where $\Gamma(z) = \int_0^\infty t^{z - 1}e^{-t} \, dt$ for every $0 < z \in \mathbb R$. If $\fnij{Y}^{[m]} = \fnij{b}W + \dij$, then
\begin{equation} \label{eq:nsst_pdf}
f_{\fnij{Y}^{[m]}}\left(\measijm \Bigm| \dij, \fnij{\bm{\theta}}\right) = \frac{1}{\fnij{b}}f_W\left(\frac{\measijm - \dij}{\fnij{b}} \Biggm| \fnij{\nu}\right),
\end{equation}
where $\fnij{\bm{\theta}} = (\fnij{\nu},\fnij{b})$.

\textbf{Evaluation.} Given an estimate $\widehat{\X}$ of $\X$, if we rotate, reflect, and/or translate the points in $\widehat{\X}$ by an equal amount, then $\widehat{\X}$ will still have the same edge lengths as $\X$. Therefore, a unique solution to the DGP cannot exist without additional constraints. Moreover, the proximity of $\widehat{\X}$ to $\X$ will need to be evaluated up to rotation, translation, and reflection. Let $\mathbf{X}$ and $\widehat{\mathbf{X}}$ be $K \times N$ matrices, where the $n$th columns of $\matr{X}$ and $\widehat{\matr{X}}$ correspond to the $n$th points in $\mathcal{X}$ and $\widehat{\X}$, respectively. We first translate the column vectors in $\mathbf{X}$ and $\widehat{\mathbf{X}}$ so that they are centered at the origin by computing $\mathbf{X}_c = \mathbf{X} - \mathbf{X}\matr{1} \big/ N$ and $\widehat{\mathbf{X}}_c = \widehat{\mathbf{X}} - \widehat{\mathbf{X}}\matr{1}\big/ N$, where $\matr{1}$ is an $N \times N$ matrix of ones.

Next, how close $\widehat{\mathcal{X}}$ is to $\mathcal{X}$ is equivalent to the solution of the orthogonal Procrustes problem (OPP) \cite{schonemann_generalized_1996},
\begin{equation} \label{eq:orth_procrustes}
    \begin{aligned}
    \min_{\matr{R}}
    \quad & \left\| \mathbf{R}\widehat{\mathbf{X}}_c - \mathbf{X}_c \right\|_F, \\
    \textrm{s.t.} \quad & \mathbf{R}^T \mathbf{R} = \matr{I},
    \end{aligned}
\end{equation}
where $\left\| \cdot \right\|_F$ is the Frobenius norm, which for a $K \times N$ matrix, $\matr{A} = (a_{ij})$, is defined as, $\left\| \matr{A} \right\|_F = \left(\sum_{k=1}^{K} \sum_{n=1}^{N} |a_{ij}|^2\right)^{1/2}$. The objective of the OPP is to determine the orthonormal matrix $\mathbf{R}$ that most closely maps $\widehat{\mathbf{X}}_c$ to $\mathbf{X}_c$. We will refer to the value obtained by solving the OPP for an estimate $\widehat{\X}$, divided by the number of points $N$, as the \emph{OPP loss} associated with $\widehat{\X}$. As our objective is to show that an ML estimate $\widehat{\X}_{\text{ML}}$ of $\X$ more closely approximates $\X$ than a non-ML estimate $\widehat{\X}$, we will compare the OPP losses of $\widehat{\X}_{\text{ML}}$ and $\widehat{\X}$.

\section{Numerical Results} \label{sec:results}

\begin{figure}[!t]
\centering
\subfloat[]{\includegraphics[width=0.9\columnwidth]{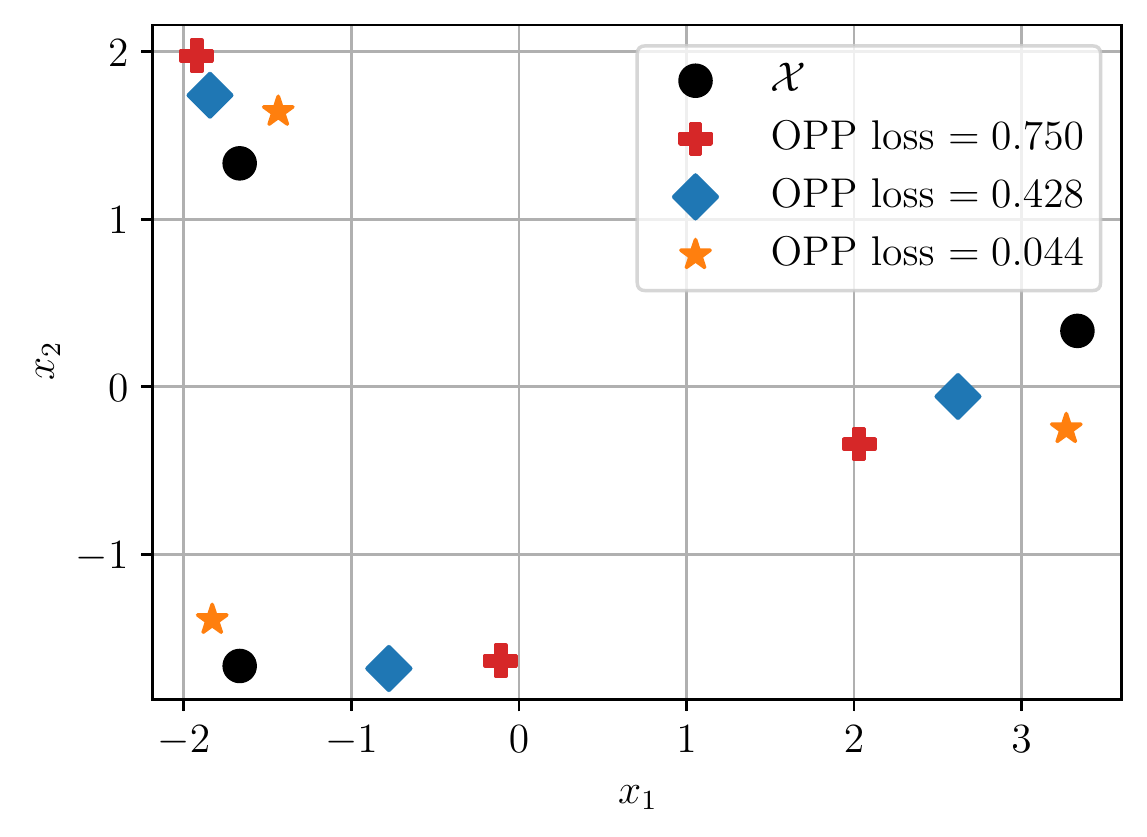}
\label{fig:tri1}}
\\
\subfloat[]{\includegraphics[width=0.9\columnwidth]{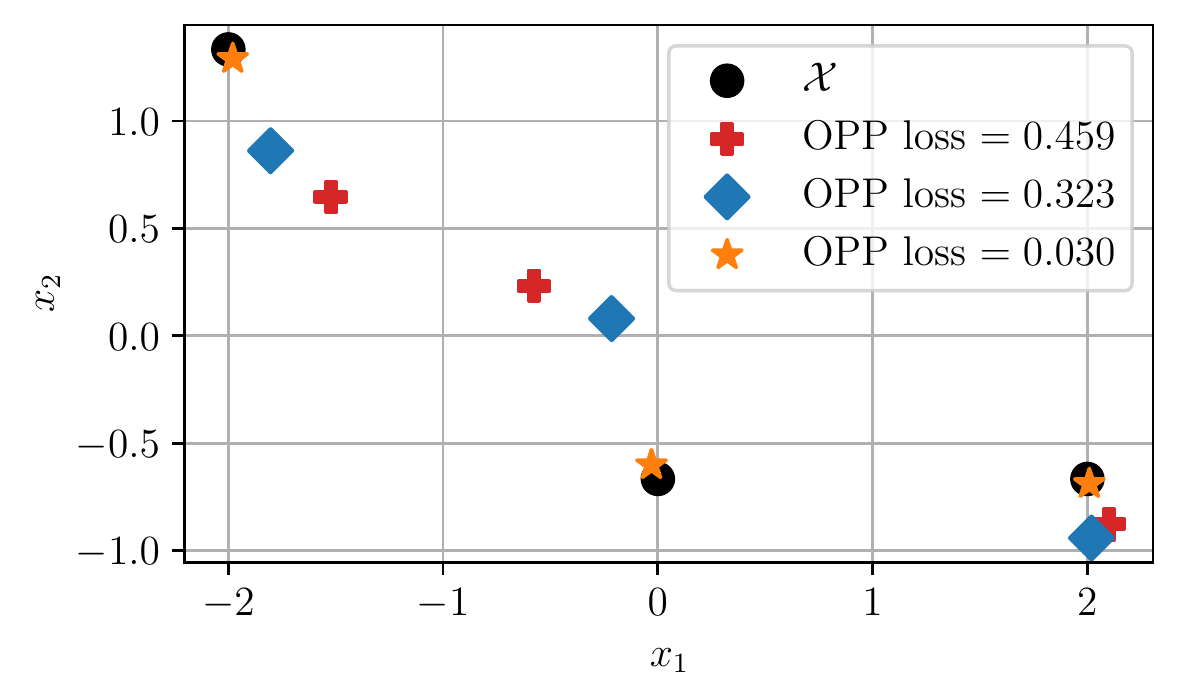}
\label{fig:tri2}}
\caption{We illustrate 2 of the 8 triangles that were considered. Ground truth structures are labeled $\X$, while example estimates, together with their associated OPP losses, are also shown. We observe that estimates with lower OPP losses more closely approximate the structure $\X$.}
\label{fig:tris}
\end{figure}

\begin{figure}[t]
\centering
\includegraphics[trim={0cm 2.5mm 0cm 0.25cm},clip,
width=0.7\columnwidth]{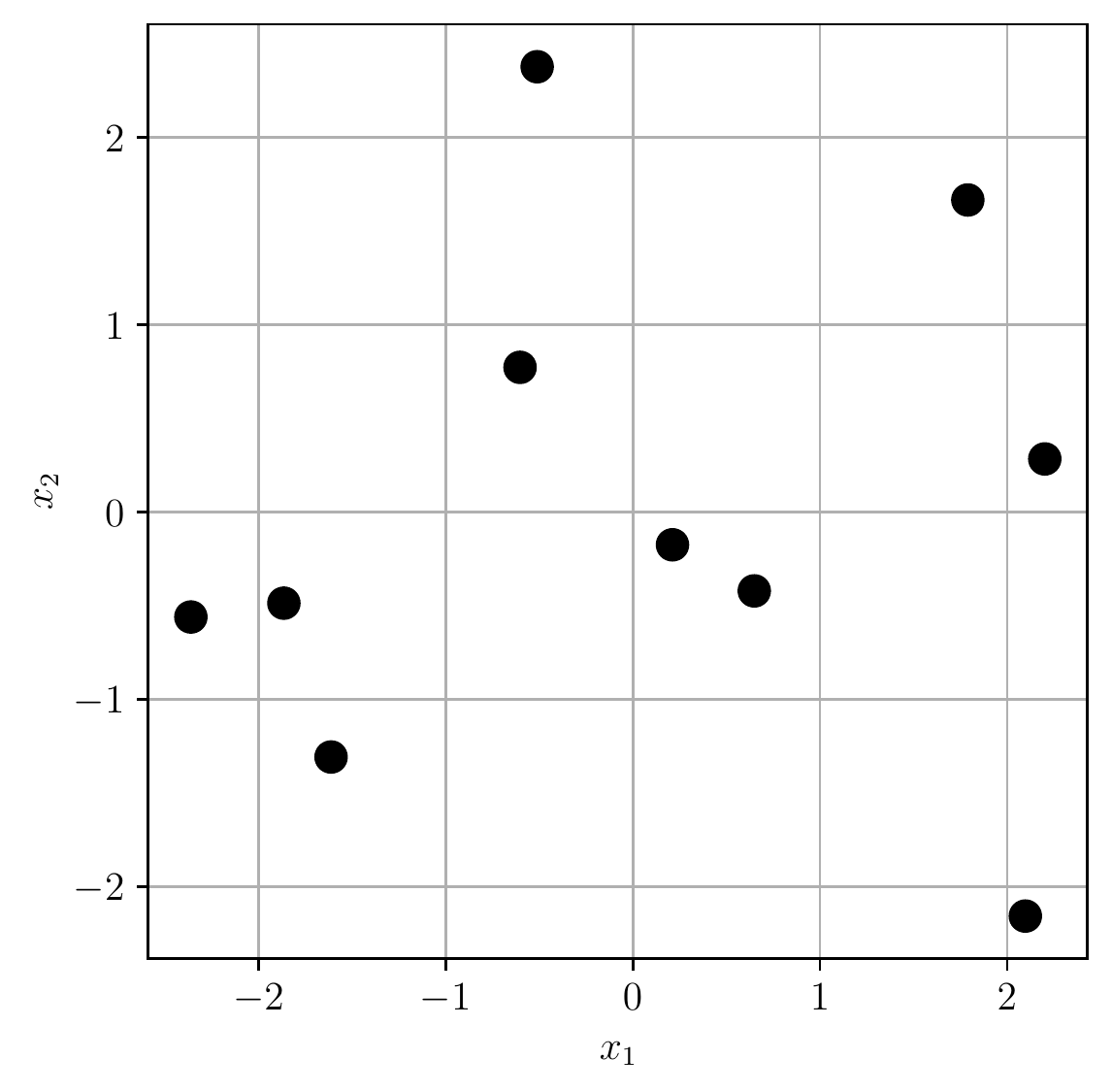} \vspace{-2mm}
\caption{One of the 30 10-point structures that were considered. Example estimates are omitted for clarity.}
\label{fig:10pts}
\end{figure}

We begin our discussion with numerical results on triangles, which are structures consisting of 3 points (see \autoref{fig:tris}). Later, we demonstrate analogous numerical results for structures with 10 points (see \autoref{fig:10pts}). In our experiments, we let $\fnij{Y}^{[m]}$ follow either a Laplace \eqref{eq:laplace_pdf} or NSST \eqref{eq:nsst_pdf} distribution. If $\fnij{Y}^{[m]}$ follows a Laplace (respectively, NSST) distribution, then the marginal pdf of $\fnij{Y}^{[m]}$ in the likelihood \eqref{eq:ll} is varied between the Laplace (respectively, NSST) pdf in the matched case and the Gaussian pdf in the mismatched case.

The variance of the noisy measurements was changed as a function of different signal-to-noise ratio (SNR) values. If the distribution from which the i.i.d. noisy measurements associated with edge $(i,j)$ were obtained has variance $\fnij{\sigma}^2$, and if we let $\sigma_\X$ be the average edge length of a structure $\X$, then $\text{SNR} = 10 \log_{10}\left(\sigma_\X^2 / \fnij{\sigma}^2 \right)$, where SNR is quantified in decibel (dB) units. We leave the case where $\fnij{\sigma}^2$ depends on edge $(i,j)$ for future work. The SNR was varied from --20 dB to 20 dB in steps of 5 dB. For each SNR value, we sample a set of noisy measurements, maximize the likelihood\footnote{An implementation of the L-BFGS algorithm \cite{nocedal_quasi-newton_2006} that is available in PyTorch \cite{paszke_pytorch_2019-1} was used to maximize the likelihood.} in \eqref{eq:ll} for the Laplace (or NSST) cases, and then report the OPP loss associated with an estimated structure $\widehat{\X}$ for each case.

\begin{figure*}[!t]
\centering
\subfloat[]{\includegraphics[width=0.66\columnwidth]{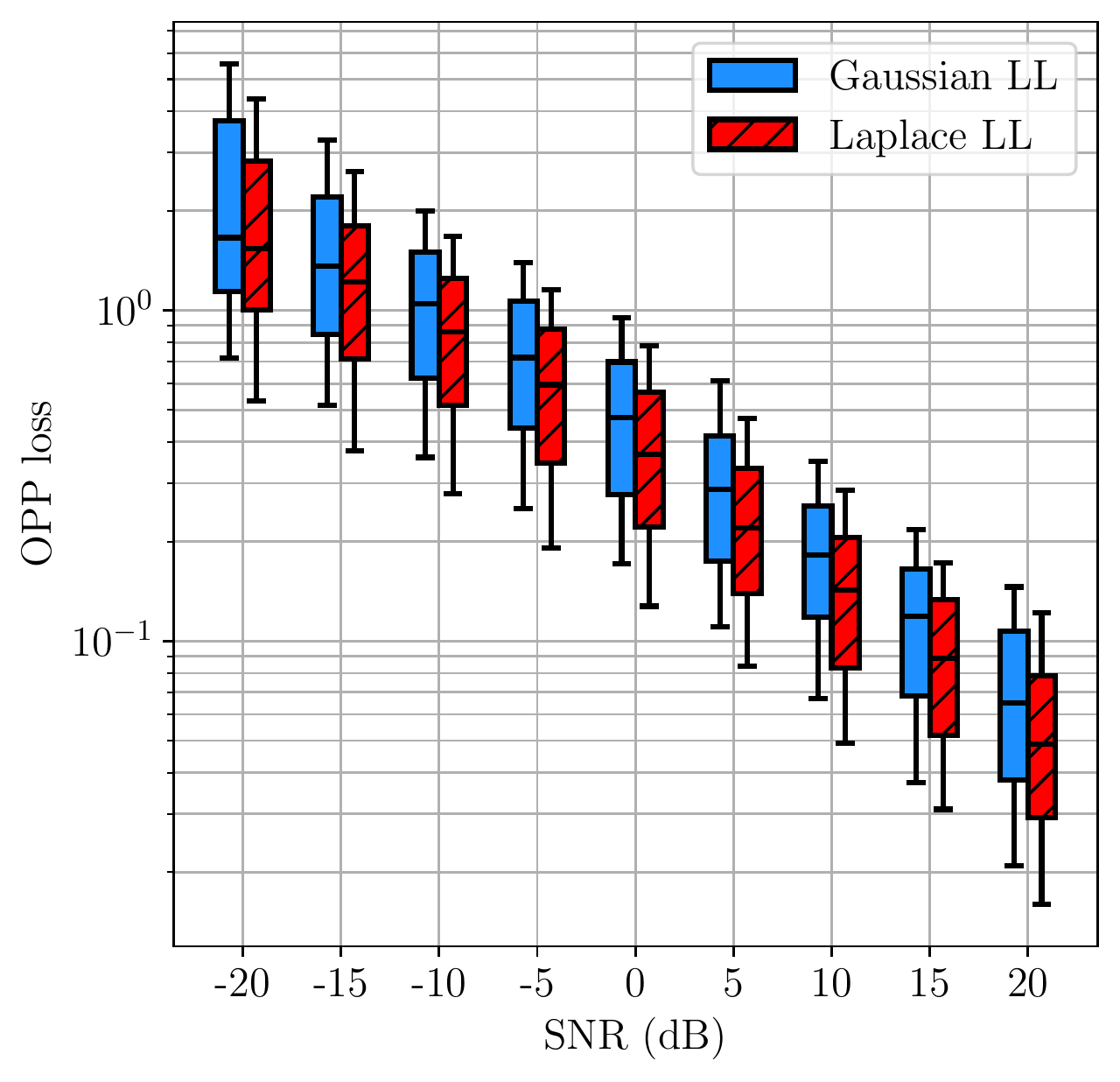}
\label{fig:3pts_M10_laplace}}
\hfil
\subfloat[]{\includegraphics[width=0.66\columnwidth]{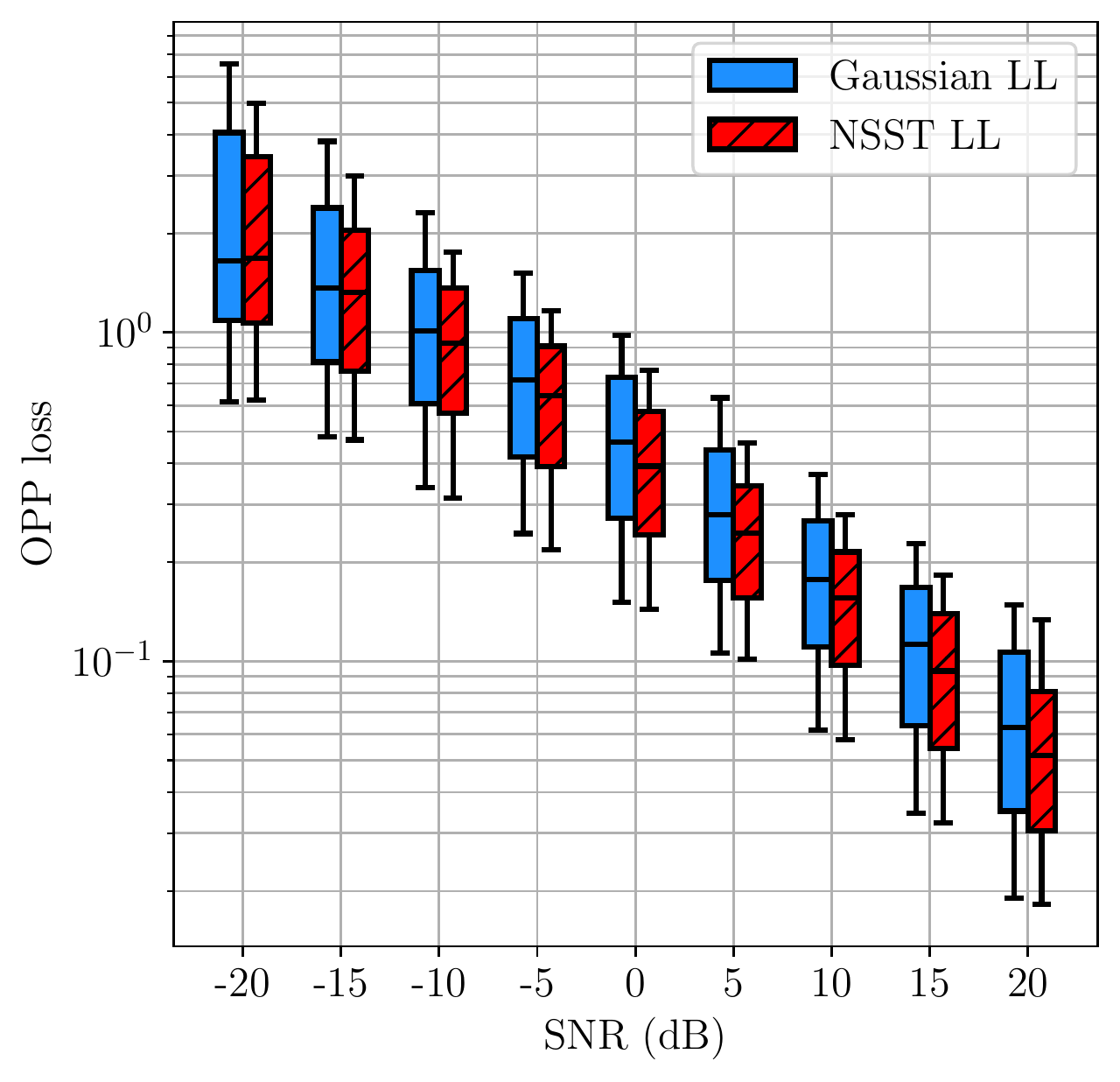}
\label{fig:3pts_M10_nsst}}
\hfil
\subfloat[]{\includegraphics[width=0.67\columnwidth]{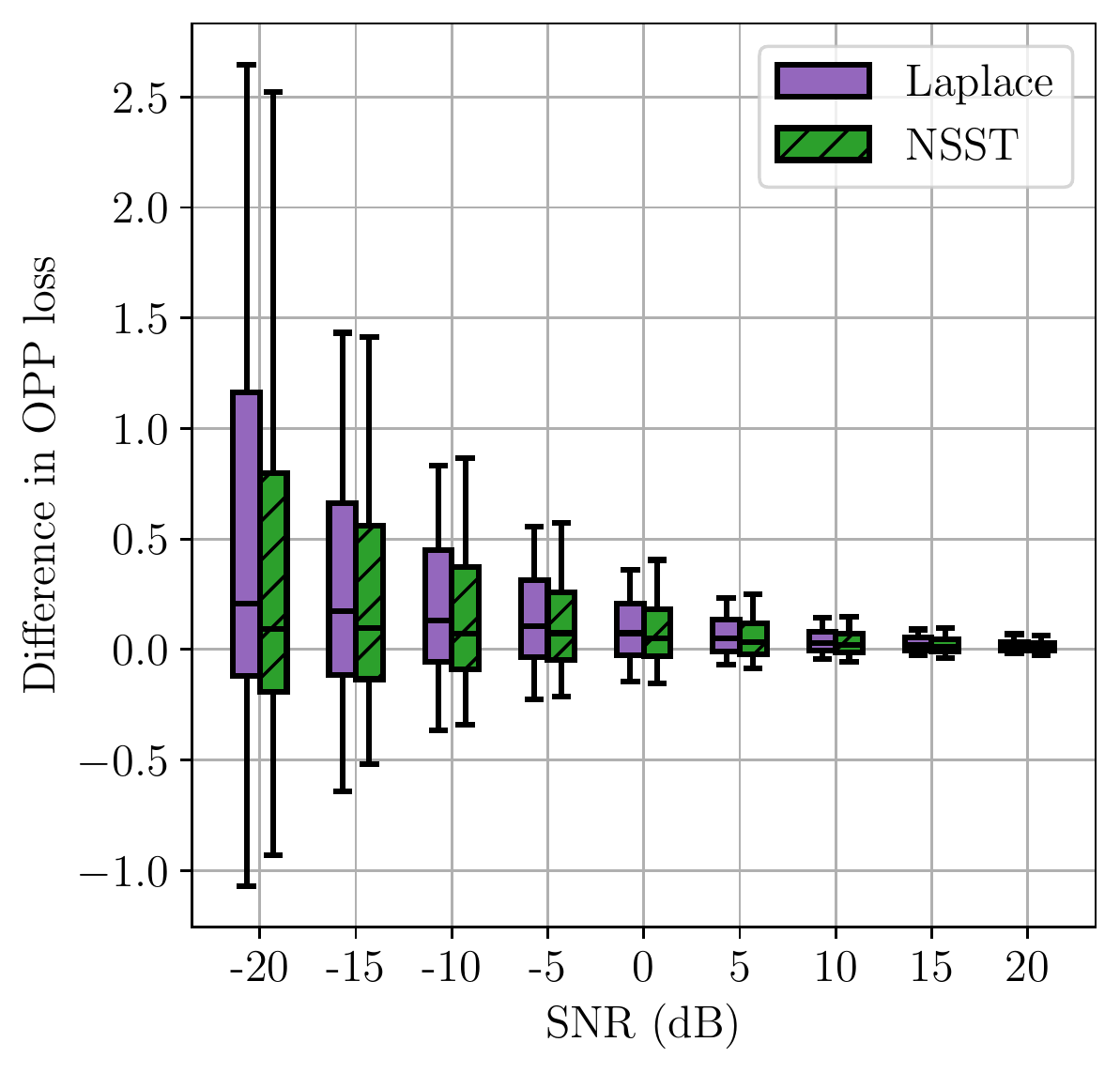}
\label{fig:3pts_M10_diff}}
\caption{Distributions of OPP losses for the 8 triangles when $\fnij{Y}^{[m]}$ follows a \protect\subref{fig:3pts_M10_laplace} Laplace or \protect\subref{fig:3pts_M10_nsst} NSST distribution, and when matched and mismatched (Gaussian) likelihood (LL) functions are used. Each box represents the percentiles (bottom to top): 10, 25, 50, 75, and 90. Distributions of pairwise differences in OPP loss when $\fnij{Y}^{[m]}$ follows a Laplace or NSST distribution are shown in \protect\subref{fig:3pts_M10_diff}. Pairwise differences are computed by subtracting the OPP loss for the matched estimate from the corresponding OPP loss for the mismatched (Gaussian LL function) estimate.}
\label{fig:3pts_M10}
\end{figure*}

\textbf{Triangles.} We first considered 8 triangles in 2D. Two of these triangles, together with example estimates and their associated OPP losses, are shown in \autoref{fig:tris}. For each SNR value, and for each triangle, we sampled $M = 10$ noisy measurements per edge 100 times. For each sample, we computed an estimate $\widehat{\X}$ using different likelihoods and then computed the OPP loss associated with $\widehat{\X}$. The distributions of these OPP losses are shown in \autoref{fig:3pts_M10_laplace} and \autoref{fig:3pts_M10_nsst}. We also computed the pairwise difference in OPP losses resulting from mismatched and matched estimation. The distributions of these differences are shown in \autoref{fig:3pts_M10_diff}.

We observe that, in general, the distribution of OPP losses associated with the likelihood that is matched to the noise distribution is shifted lower than the distribution associated with the likelihood that is not matched to the noise distribution. Moreover, in several instances, the increase in median OPP loss due to a 3--4 dB decrease in SNR can approximately be compensated for by choosing to use the likelihood that is matched to the noise distribution.

\begin{figure}[!t]
\centering
\subfloat[]{\includegraphics[trim={2.65mm 0mm 0mm 2.5mm},clip,
width=0.532\columnwidth]{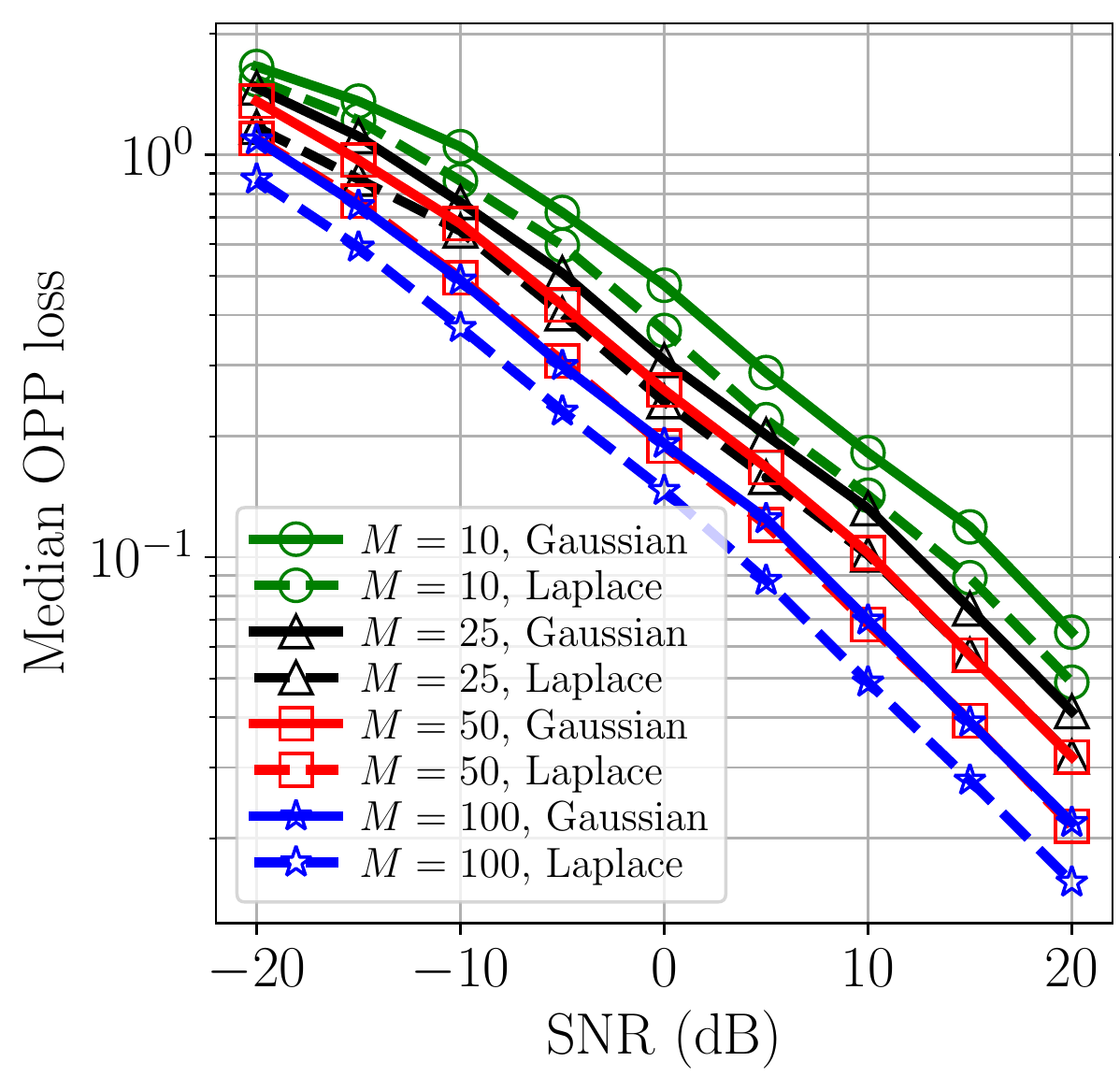}
\label{fig:3pts_Ms_laplace}}
\hspace{-2.3mm}
\subfloat[]{\includegraphics[trim={0cm 1.2mm 2.3mm 2mm},clip,
width=0.441\columnwidth]{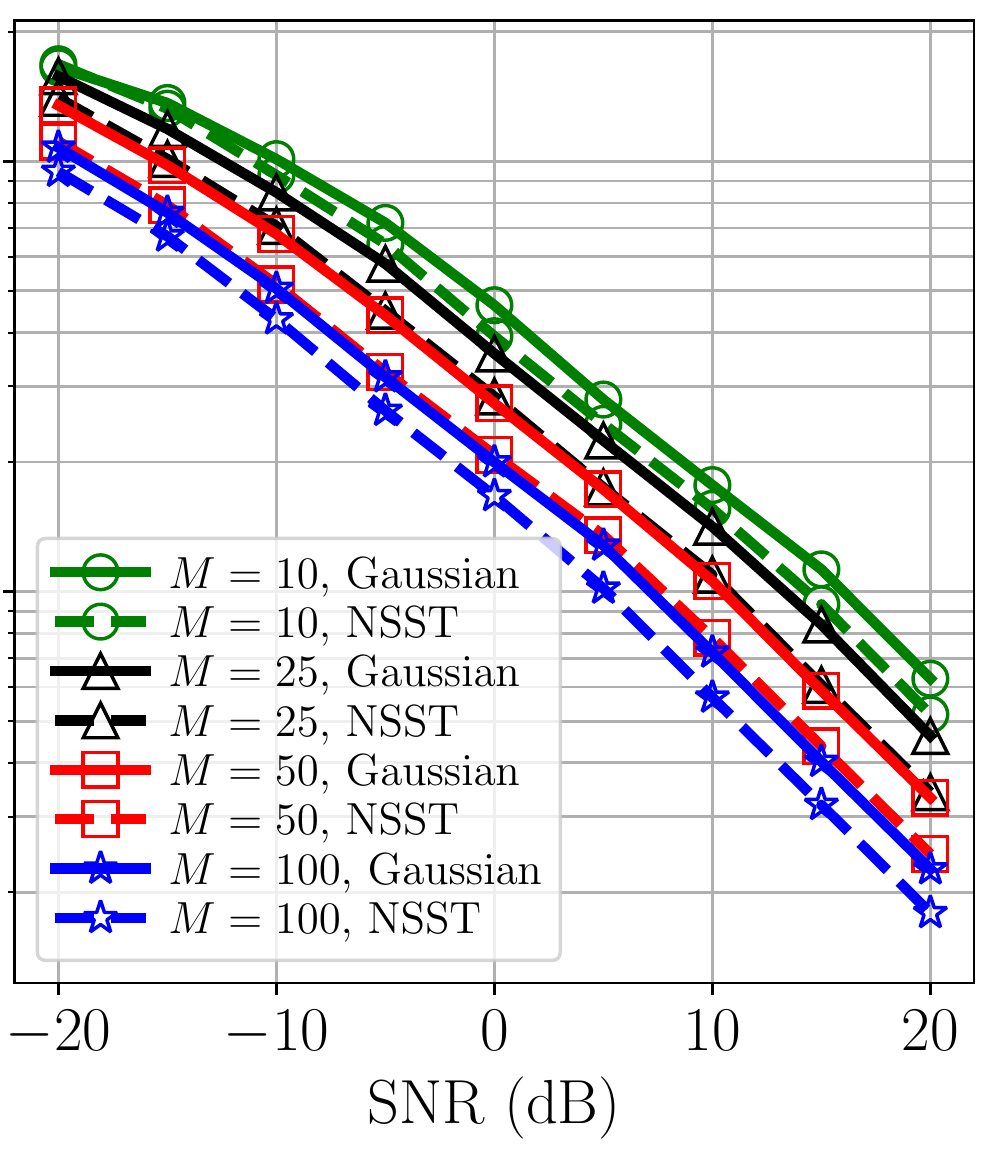}
\label{fig:3pts_Ms_nsst}}
\caption{Median OPP losses for the 8 triangles and for different $M$ values when $\fnij{Y}^{[m]}$ follows a \protect\subref{fig:3pts_Ms_laplace} Laplace or \protect\subref{fig:3pts_Ms_nsst} NSST distribution, and when matched or mismatched (Gaussian) likelihood functions are used.}
\label{fig:3pts_Ms}
\end{figure}

We then repeated this experiment for different values of $M$, the number of noisy measurements per edge. The median OPP losses for $M = 10,25,50,100$ are shown in \autoref{fig:3pts_Ms}. We observe that, in general, maximizing the likelihood that is matched to the noise distribution offers an improvement over mismatched DGP estimation comparable to the improvement obtained by doubling the number of noisy measurements per edge.

\begin{figure*}[!t]
\centering
\subfloat[]{\includegraphics[width=0.66\columnwidth]{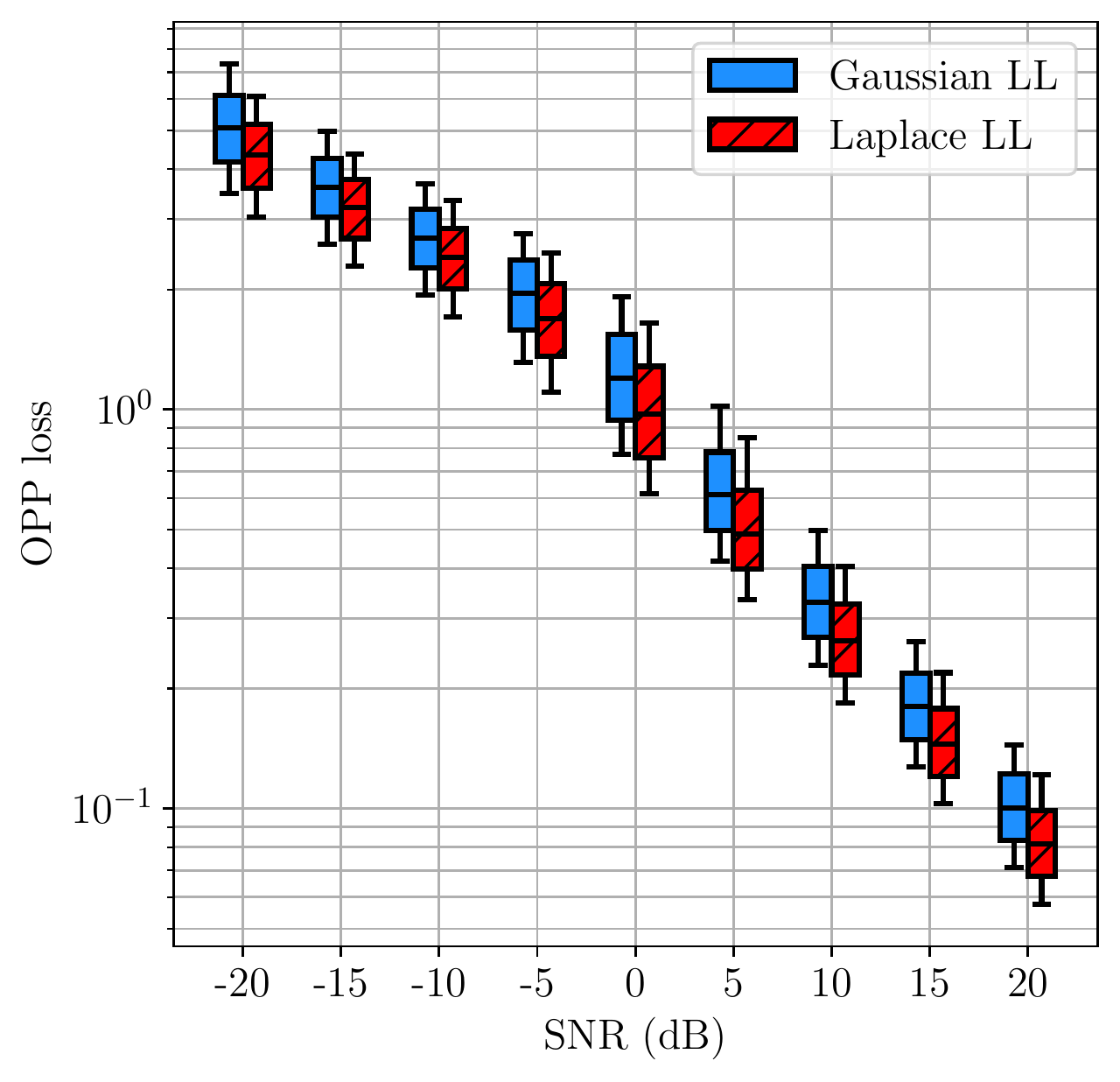}
\label{fig:10pts_M10_laplace}}
\hfil
\subfloat[]{\includegraphics[width=0.66\columnwidth]{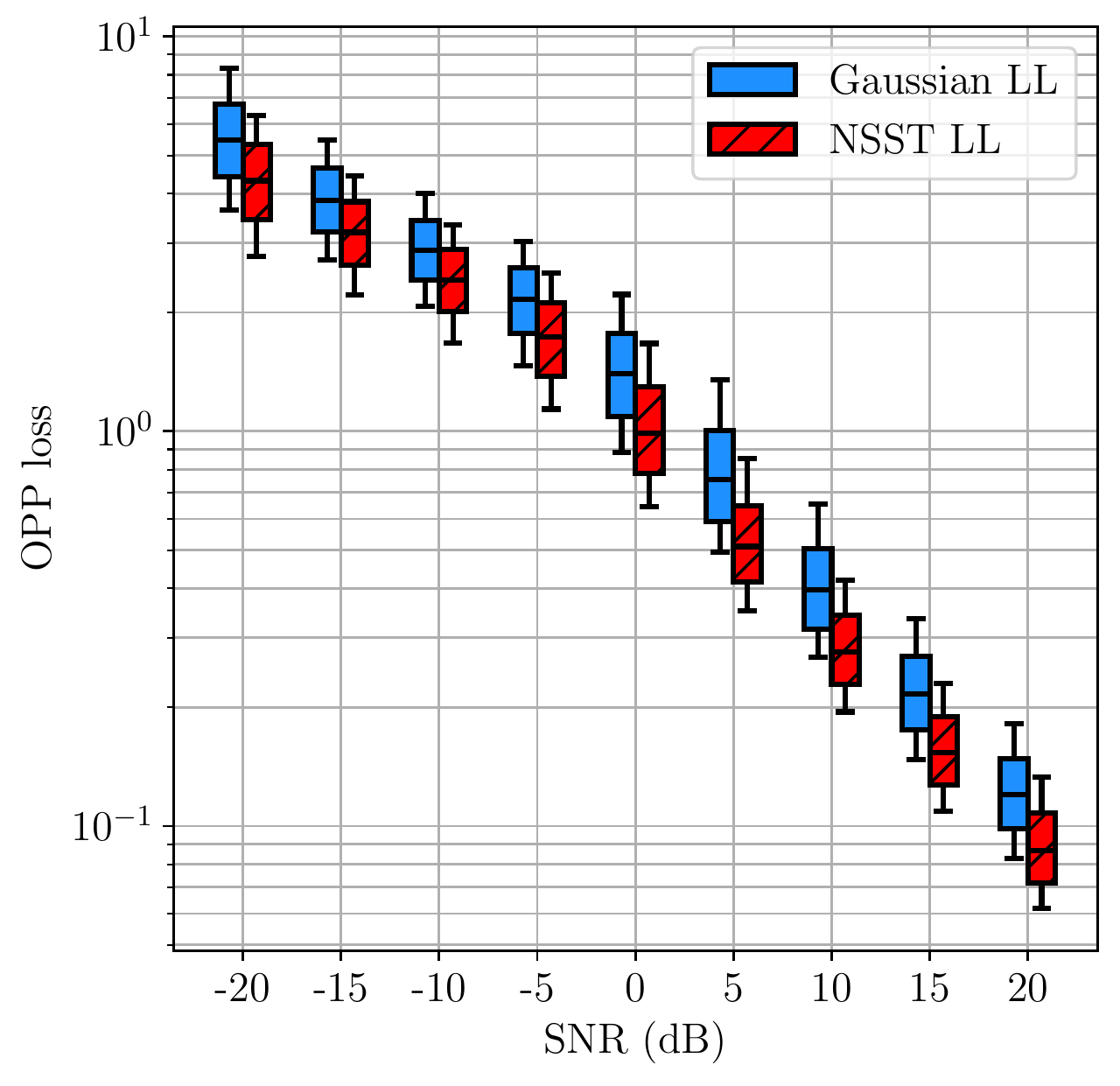}
\label{fig:10pts_M10_nsst}}
\hfil
\subfloat[]{\includegraphics[width=0.67\columnwidth]{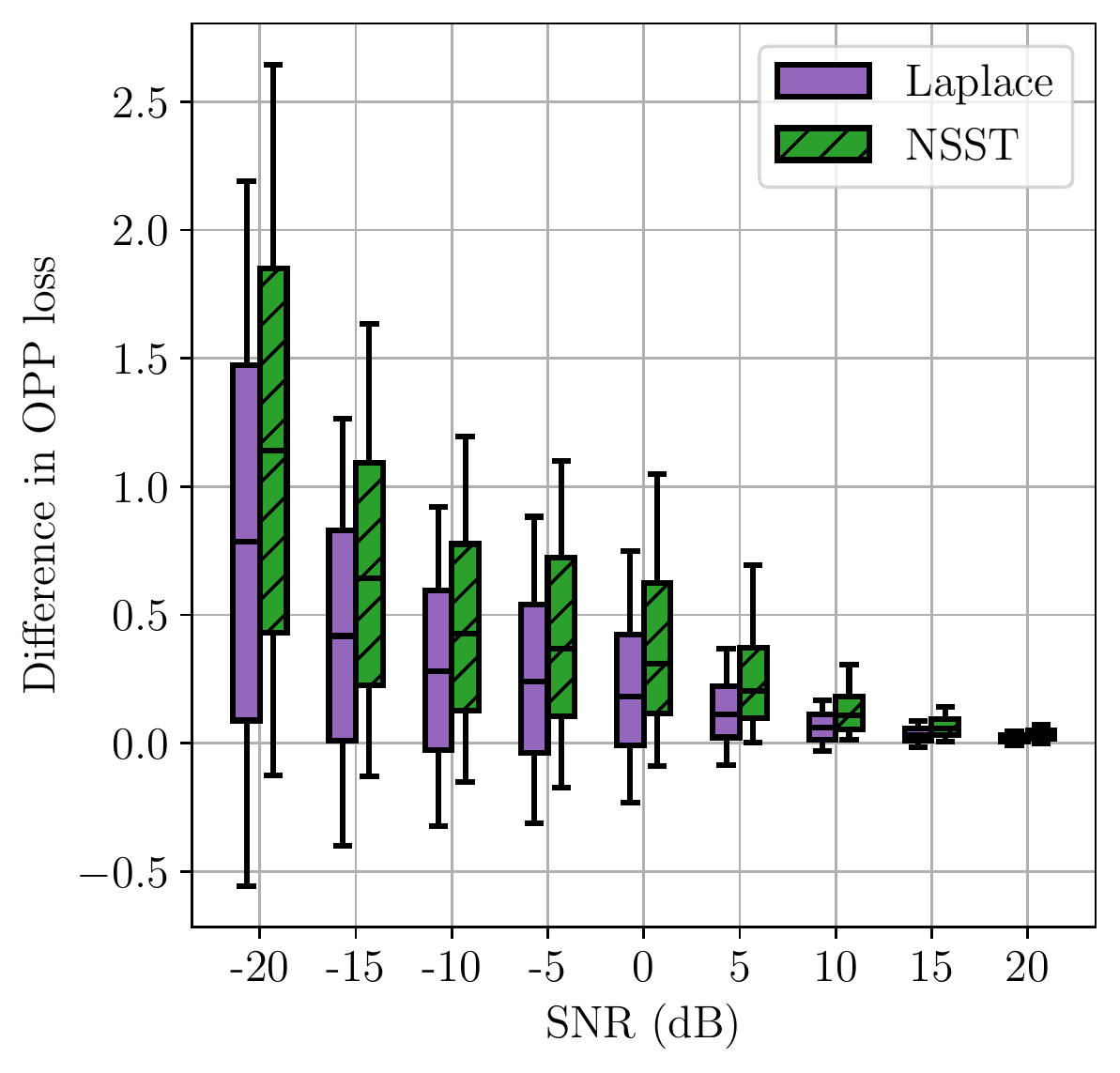}
\label{fig:10pts_M10_diff}}
\caption{Distributions of OPP losses for the 30 10-point structures when $\fnij{Y}^{[m]}$ follows a \protect\subref{fig:10pts_M10_laplace} Laplace or \protect\subref{fig:10pts_M10_nsst} NSST distribution, and when matched and mismatched (Gaussian) likelihood (LL) functions are used. Percentiles are marked as in \autoref{fig:3pts_M10}. Distributions of pairwise differences in OPP loss when $\fnij{Y}^{[m]}$ follows a Laplace or NSST distribution are shown in \protect\subref{fig:10pts_M10_diff}. Pairwise differences are computed by subtracting the OPP loss for the matched estimate from the corresponding OPP loss for the mismatched (Gaussian LL function) estimate.}
\label{fig:10pts_M10}
\end{figure*}

\textbf{10 points.} To further support our results, we repeated our experiments for 30 10-point structures in 2D. One of these structures is shown in \autoref{fig:10pts}. The distributions of OPP losses are shown in \autoref{fig:10pts_M10_laplace} and \autoref{fig:10pts_M10_nsst}, and the distributions of pairwise differences in OPP losses for the mismatched and matched cases are shown in \autoref{fig:10pts_M10_diff}. Similar to the case of triangles, we observe that, in general, the distribution of OPP losses associated with the likelihood that is matched to the noise distribution is lower than the distribution associated with the likelihood that is not matched to the noise distribution.

\section{Conclusions and Future Work} \label{sec:conclusion}

In this paper, we have explored the impact of matching the likelihood function to the noise distribution on estimation accuracy in the context of the distance geometry problem (DGP). We have empirically found that matching the likelihood function to the noise distribution can offer a 3--4 dB compensation for the loss in estimation accuracy due to noisier measurements and smaller sample sizes. Moreover, matched estimation can be helpful in these scenarios, which often arise in practice.

There are several directions for future work that can be explored. For example, in practice, measurements may not be available for some edges, so the case where the numbers of edge measurements are unequal may also be considered. We may also explore the application of our approach to real-world measurements. Finally, a theoretical evaluation of the mismatch between the likelihood function and the noise distribution would be an insightful endeavor.

\section*{Acknowledgments}

We thank Vaibhav Choudhary and Nikhil Gopalkrishnan for informative discussions.

\printbibliography

@book{crippen_distance_1988,
  title = {Distance Geometry and Molecular Conformation},
  author = {Crippen, G. M. and Havel, Timothy F.},
  date = {1988},
  eprint = {5CPwAAAAMAAJ},
  eprinttype = {googlebooks},
  publisher = {{Research Studies Press}},
  isbn = {978-0-86380-073-3},
  langid = {english}
}

@inproceedings{eren_rigidity_2004,
  title = {Rigidity, Computation, and Randomization in Network Localization},
  booktitle = {{{IEEE INFOCOM}}},
  author = {Eren, T. and Goldenberg, O.K. and Whiteley, W. and Yang, Y.R. and Morse, A.S. and Anderson, B.D.O. and Belhumeur, P.N.},
  date = {2004-03},
  volume = {4},
  pages = {2673--2684},
  issn = {0743-166X},
  doi = {10.1109/INFCOM.2004.1354686},
  eventtitle = {{{IEEE INFOCOM}}}
}

@article{giamou_convex_2022,
  title = {Convex Iteration for Distance-Geometric Inverse Kinematics},
  author = {Giamou, Matthew and Marić, Filip and Rosen, David M. and Peretroukhin, Valentin and Roy, Nicholas and Petrović, Ivan and Kelly, Jonathan},
  date = {2022-04},
  journaltitle = {IEEE Robot. Autom. Lett.},
  volume = {7},
  number = {2},
  pages = {1952--1959},
  issn = {2377-3766},
  doi = {10.1109/LRA.2022.3141763},
  eventtitle = {{{IEEE Robotics}} and {{Automation Letters}}}
}

@article{gopalkrishnan_dna_2020-1,
  title={A {DNA} nanoscope that identifies and precisely localizes over a hundred unique molecular features with nanometer accuracy},
  author={Gopalkrishnan, Nikhil and Punthambaker, Sukanya and Schaus, Thomas E and Church, George M and Yin, Peng},
  journal={bioRxiv},
  year={2020},
}

@article{havel_evaluation_1991,
  title = {An Evaluation of Computational Strategies for Use in the Determination of Protein Structure from Distance Constraints Obtained by Nuclear Magnetic Resonance},
  author = {Havel, T. F.},
  date = {1991-01},
  journaltitle = {Prog. Biophys. Mol. Biol.},
  volume = {56},
  number = {1},
  pages = {43--78},
  issn = {0079-6107},
  doi = {10.1016/0079-6107(91)90007-F},
  url = {https://www.sciencedirect.com/science/article/pii/007961079190007F},
  urldate = {2021-09-04},
  langid = {english}
}

@article{hendrickson_molecule_1995,
  title = {The Molecule Problem: {{Exploiting}} Structure in Global Optimization},
  shorttitle = {The Molecule Problem},
  author = {Hendrickson, Bruce},
  date = {1995-11},
  journaltitle = {SIAM J. Optim.},
  volume = {5},
  number = {4},
  pages = {835--857},
  publisher = {{Society for Industrial and Applied Mathematics}},
  issn = {1052-6234},
  doi = {10.1137/0805040},
  url = {https://epubs-siam-org.prox.lib.ncsu.edu/doi/10.1137/0805040},
  urldate = {2021-09-04}
}

@incollection{lavor_overview_2009,
  title = {An Overview of Distinct Approaches for the Molecular Distance Geometry Problem},
  booktitle = {Encyclopedia of {{Optimization}}},
  author = {Lavor, Carlile and Liberti, Leo and Maculan, Nelson},
  editor = {Floudas, Christodoulos A. and Pardalos, Panos M.},
  date = {2009},
  pages = {2304--2311},
  publisher = {{Springer US}},
  location = {{Boston, MA}},
  doi = {10.1007/978-0-387-74759-0_400},
  url = {https://doi.org/10.1007/978-0-387-74759-0_400},
  urldate = {2021-09-04},
  isbn = {978-0-387-74759-0},
  langid = {english}
}

@incollection{liberti_distance_2017-1,
  title = {The Distance Geometry Problem},
  booktitle = {Euclidean {{Distance Geometry}}: {{An Introduction}}},
  author = {Liberti, Leo and Lavor, Carlile},
  editor = {Liberti, Leo and Lavor, Carlile},
  date = {2017},
  pages = {9--18},
  publisher = {{Springer International Publishing}},
  location = {{Cham}},
  doi = {10.1007/978-3-319-60792-4_2},
  url = {https://doi.org/10.1007/978-3-319-60792-4_2},
  urldate = {2022-03-28},
  isbn = {978-3-319-60792-4},
  langid = {english}
}

@incollection{liberti_open_2018,
  title = {Open Research Areas in Distance Geometry},
  booktitle = {Open Problems in Optimization and Data Analysis},
  author = {Liberti, Leo and Lavor, Carlile},
  date = {2018},
  pages = {183--223},
  publisher = {{Springer}}
}

@article{luo_least-squares_2011,
  title = {Least-Squares Approximations in Geometric Buildup for Solving Distance Geometry Problems},
  author = {Luo, Xin-Long and Wu, Zhi-Jun},
  date = {2011-06-01},
  journaltitle = {J. Optim. Theory Appl.},
  volume = {149},
  number = {3},
  pages = {580--598},
  issn = {1573-2878},
  doi = {10.1007/s10957-011-9806-6},
  url = {https://doi.org/10.1007/s10957-011-9806-6},
  urldate = {2022-04-08},
  langid = {english}
}

@article{more_global_1997-2,
  title = {Global Continuation for Distance Geometry Problems},
  author = {Moré, Jorge J. and Wu, Zhijun},
  date = {1997-08-01},
  journaltitle = {SIAM J. Optim.},
  volume = {7},
  number = {3},
  pages = {814--836},
  publisher = {{Society for Industrial and Applied Mathematics}},
  issn = {1052-6234},
  doi = {10.1137/S1052623495283024},
  url = {https://epubs-siam-org.prox.lib.ncsu.edu/doi/10.1137/S1052623495283024},
  urldate = {2021-09-04}
}

@incollection{nocedal_quasi-newton_2006,
  title = {Quasi-{{Newton}} Methods},
  booktitle = {Numerical {{Optimization}}},
  editor = {Nocedal, Jorge and Wright, Stephen J.},
  date = {2006},
  series = {Springer {{Series}} in {{Operations Research}} and {{Financial Engineering}}},
  pages = {135--163},
  publisher = {{Springer}},
  location = {{New York, NY}},
  doi = {10.1007/978-0-387-40065-5_6},
  url = {https://doi.org/10.1007/978-0-387-40065-5_6},
  urldate = {2021-11-17},
  isbn = {978-0-387-40065-5},
  langid = {english}
}

@article{paszke_pytorch_2019-1,
  title={Pytorch: An imperative style, high-performance deep learning library},
  author={Paszke, Adam and Gross, Sam and Massa, Francisco and Lerer, Adam and Bradbury, James and Chanan, Gregory and Killeen, Trevor and Lin, Zeming and Gimelshein, Natalia and Antiga, Luca and others},
  journal={Advances in Neural Information Processing Systems},
  volume={32},
  year={2019},
}

@article{schonemann_generalized_1996,
  title = {A Generalized Solution of the Orthogonal Procrustes Problem},
  author = {Schönemann, Peter H.},
  date = {1996-03},
  journaltitle = {Psychometrika},
  volume = {31},
  number = {1},
  pages = {1--10},
  issn = {1860-0980},
  doi = {10.1007/BF02289451},
  url = {https://doi.org/10.1007/BF02289451},
  urldate = {2021-11-19},
  langid = {english}
}

@article{souza_hyperbolic_2011,
  title = {Hyperbolic Smoothing and Penalty Techniques Applied to Molecular Structure Determination},
  author = {Souza, Michael and Xavier, Adilson Elias and Lavor, Carlile and Maculan, Nelson},
  date = {2011-11-01},
  journaltitle = {Operations Research Letters},
  volume = {39},
  number = {6},
  pages = {461--465},
  issn = {0167-6377},
  doi = {10.1016/j.orl.2011.07.007},
  url = {https://www.sciencedirect.com/science/article/pii/S0167637711000848},
  urldate = {2022-04-08},
  langid = {english}
}

@article{souza_solving_2013,
  title = {Solving the Molecular Distance Geometry Problem with Inaccurate Distance Data},
  author = {Souza, Michael and Lavor, Carlile and Muritiba, Albert and Maculan, Nelson},
  date = {2013-06-28},
  journaltitle = {BMC Bioinform.},
  volume = {14},
  eprint = {23901894},
  eprinttype = {pmid},
  pages = {S7},
  issn = {1471-2105},
  doi = {10.1186/1471-2105-14-S9-S7},
  url = {https://www.ncbi.nlm.nih.gov/pmc/articles/PMC3698034/},
  urldate = {2021-11-15},
  issue = {Suppl 9},
  pmcid = {PMC3698034}
}

@incollection{voller_distance_2013,
  title = {Distance Geometry Methods for Protein Structure Determination},
  booktitle = {Distance {{Geometry}}: {{Theory}}, {{Methods}}, and {{Applications}}},
  author = {Voller, Zachary and Wu, Zhijun},
  editor = {Mucherino, Antonio and Lavor, Carlile and Liberti, Leo and Maculan, Nelson},
  date = {2013},
  pages = {139--159},
  publisher = {{Springer}},
  location = {{New York, NY}},
  doi = {10.1007/978-1-4614-5128-0_8},
  url = {https://doi.org/10.1007/978-1-4614-5128-0_8},
  urldate = {2022-03-28},
  isbn = {978-1-4614-5128-0},
  langid = {english}
}
\end{document}